\documentclass[]{spie}  %>>> use for US letter paper
\usepackage{amsmath,amsfonts,amssymb}
\usepackage{graphicx}
\usepackage[colorlinks=true, allcolors=blue]{hyperref}

\usepackage{aas_macros}

\title{Second generation spectroscopic instrumentation for the STELLA robotic observatory}

\author[a]{Michael Weber}
\author[a]{Manfred Woche}
\author[a]{Klaus G. Strassmeier}
\author[a]{Ilya Ilyin}
\author[a]{Arto J\"arvinen}
\affil[a]{Leibniz-Institute for Astrophysics Potsdam (AIP), An der Sternwarte 16, D--14482 Potsdam, Germany}

\authorinfo{Send correspondence to M. Weber: E-mail: mweber@aip.de}

\newcommand{\ms}{m\,s$^{-1}$}

\def\degr{\hbox{$^\circ$}}

\begin{document} 
\maketitle

\begin{abstract}
The current STELLA \'Echelle spectrograph (SES), which records 390nm to 870nm in one shot at a spectral resolution of 55000, will be replaced by a suite of specialized spectrographs in three spectral bands. The UV will be covered by a newly designed H\&K spectrograph covering 380\,nm to 470\,nm (SES-H\&K), the visual band (470\,nm - 690\,nm) will be covered by SES-VIS, which is a vacuum-stabilized spectrograph designed for high radial-velocity accuracy, and the NIR will be covered by the current SES spectrograph from 690\,nm  to 1050\,nm. 
In order to improve the UV transmission, and to accommodate three different fibre-feeds, the prime focus corrector of the telescope will be refurbished, leading to an optical system with the f/2 1200mm spherical primary, a 4-lens collimator with 2" aperture, atmospheric dispersion corrector (ADC),  and two dichroic beam splitters, feeding 3 separate fibre feeds for the three bands. 
The newly designed H\&K spectrograph will be an \'Echelle spectrograph, based on a R4-grating with 41.6 l/mm and 110mmx420mm, using a f/5 camera and the cross-disperser in double pass (as in TRAFICOS, MIKE, KPF), using 21 spectral orders.
The spectral resolution of all three spectrographs will be comparable to the current SES's 55000. 
\end{abstract}

\keywords{instrumentation: spectrographs -- observatory: robotic}

\section{INTRODUCTION}
\label{sec:intro}  % \label{} allows reference to this section

STELLA \cite{2004AN....325..527S} (short for STELLar Activity) is a fully autonomous observatory located at Teide observatory on Tenerife, Spain. The Teide observatory is located at Iza{\~n}a peak, longitude $16^\circ 30'35"$W and at latitude $28^{\circ} 18'00"$N at an altitude of 2390m above sea level. STELLA consists of two independent, 1.2m telescopes, each of them serving a single instrument to avoid the necessity of instrument change. Both telescopes have been manufactured by Halfmann Teleskoptechnik in Augsburg, Germany and are modern alt-azimuthal telescopes with a rather high slewing speed of 10\,$^{\circ}$\!/s. 
The observatory started operations in 2006 with SES fed from the STELLA-I telescope. In early 2012,  STELLA-I was equipped with an imager, and the fiber feed moved to STELLA-II. Since then a fiber in the prime focus of the then newly commissioned STELLA-II telescope is used to feed SES. 

\begin{figure}[ht]
\centering
\includegraphics[width=\textwidth]{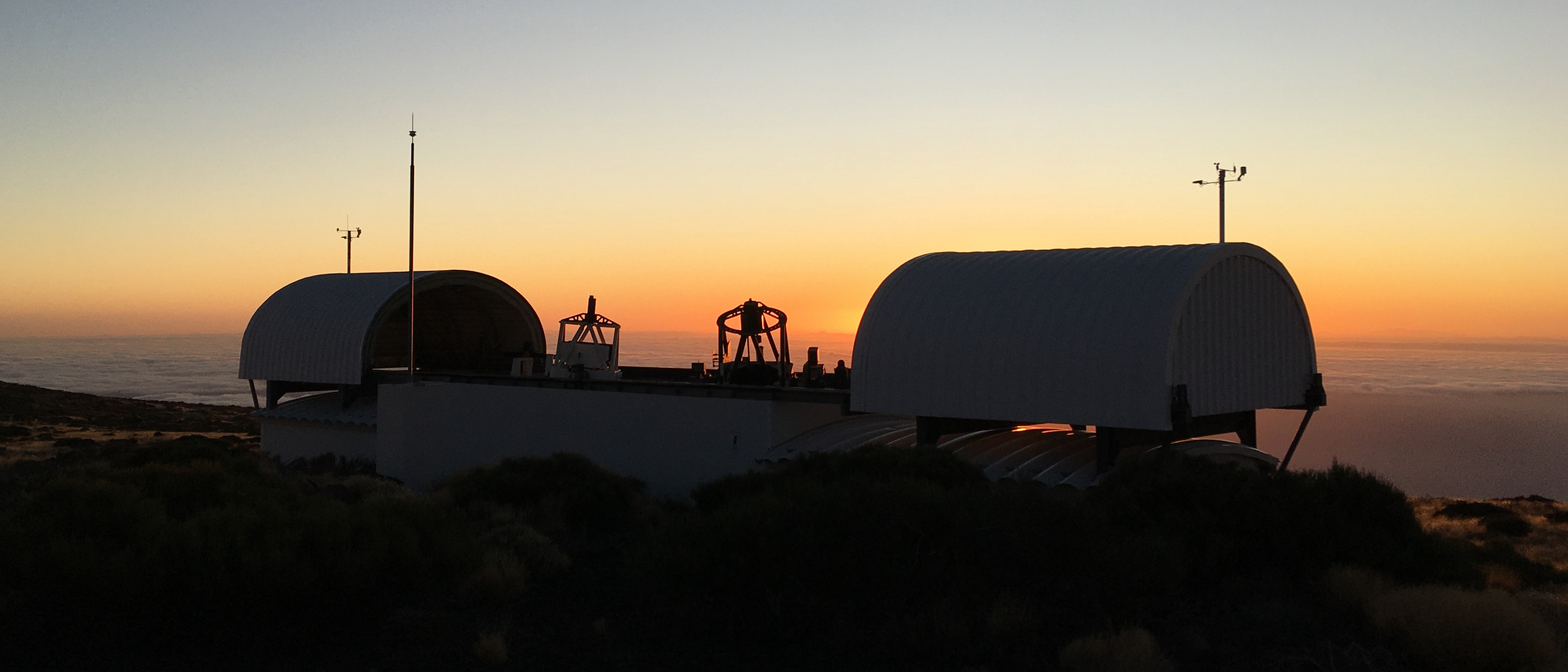}
\caption{\label{fig:observatory}
View of the STELLA observatory just before sunrise, the roof is still open for calibration purposes. The STELLA-I telescope is located at the right (West) side of the building and is recording sky flat fields, the STELLA-II telescope is located at the left (East side), recording sky reference spectra.}
\end{figure}

\section{MOTIVATION FOR THE UPGRADE}

The main objective of the spectrograph (and the whole observatory) is stellar activity research. Since we started recording our first spectra in 2005 we were unhappy with the blue throughput of the spectrograph-telescope system, and since then have been trying to overcome bottlenecks. That resulted in a spectrograph upgrade that improved the situation, but due to the long fiber from the telescope to the spectrograph room (20\,m) could not completely overcome the shortcomings. A radial velocity precision of better than the current 30\ms\ (e.~g.~Weber \& Strassmeier 2011\cite{2011A&A...531A..89W}) is also increasingly expected, but is impossible to obtain with incremental upgrades of the current system.

We therefore decided to install a new vacuum-enclosed visual spectrograph as the new workhorse instrument. Limiting the wavelength range to the visual part made the footprint of the spectrograph small enough to fit in a quasi-standard 500\,mm vacuum tube and still match the spectral resolution of the previous system (see Sect.~\ref{sec:sesvis}). The spectral region with wavelength shorter than the visual will be covered by a dedicated UV-spectrograph which is to be mounted close to the telescope to keep the fiber length below 5\,m (see Sect.~\ref{sec:seshk}). The NIR will be covered by the existing spectrograph up to the cut-off of the existing CCD (about 1\,$\mu$m). Plans to cover the NIR up to 1.1$\mu$m exist, but sourcing the right detector system remains a problem for now (see Sect.~\ref{sec:sesnir}).

To feed three spectrographs at the same time needs a modified fiber-feed unit at the telescope (Sect~\ref{sec:stella2corr}). As our original prime-focus corrector was built as a plug-in replacement for the original STELLA-I fiber-feed, it provided a f/8 output. That made the system quite long, and makes it impossible to insert beam splitters in the current unit. 

Mechanically the current fiber feed was based on a commercial cage system. That made the setup very flexible and allowed for easy experiments using off-the-shelf components. But the drawback was, that the structure including the correction optics were rigidly connected to the spider structure, and alignment could only be done by tip/tilt of the spider. This resulted in a less-than-optimal PSF which was partly bypassed by using a bigger fiber, but a slight loss of radial velocity precision after changing from STELLA-I to STELLA-II was probably caused by the uneven fiber illumination caused by the imperfect alignment.

\section{THE NEW STELLA-II CORRECTOR}
\label{sec:stella2corr}

\begin{figure}[hb]
\begin{minipage}[c]{.68\textwidth}
\includegraphics[width=\textwidth]{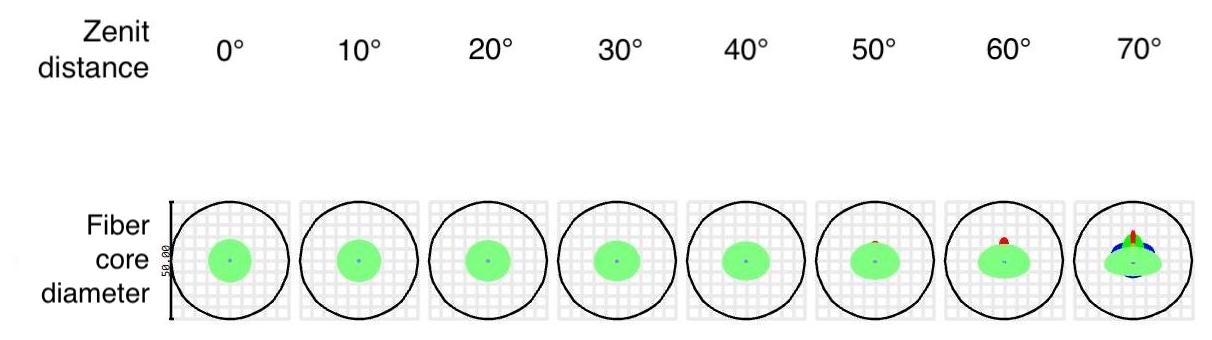}%
\end{minipage}\hfill
\begin{minipage}[t]{.3\textwidth}
\caption{\label{fig:adc}
Performance of the ADC: Simulated spot sizes and positions for the wavelength range of all three spectrographs, compared for Zenit distances of 0\degr to 70\degr. }
\end{minipage}
\end{figure}

The STELLA-II prime focus corrector corrects for the spherical aberration of the 1.2\,m spherical main mirror of the telescope. The current corrector was designed to create a f/8 beam similar to the STELLA-I telescope, so we could keep the same micro-lens fiber coupling as was used there. Due to the tight space constraints, we used a tilted pinhole mirror to use the spilled light for guiding. Four ADC prisms for four different Zenit angles were put into a small optics wheel which was also used to couple calibration light into the fiber. The fiber, the optics wheel and the guiding camera were mounted on a slider to adjust for focus changes due to thermal expansion of the telescope tube \cite{2016SPIE.9910E..0NW}.

Because of the need for an update of the corrector to allow for two beamsplitters to feed the three spectrographs with their corresponding wavelength range, we took the opportunity to redesign the unit for optimal blue transparency and to make it more compact. At the same time we want to make guiding more efficient by using a beamsplitter that also allows to control the illumination of the fiber. 

\begin{figure}[ht]
\centering
\includegraphics[width=\textwidth]{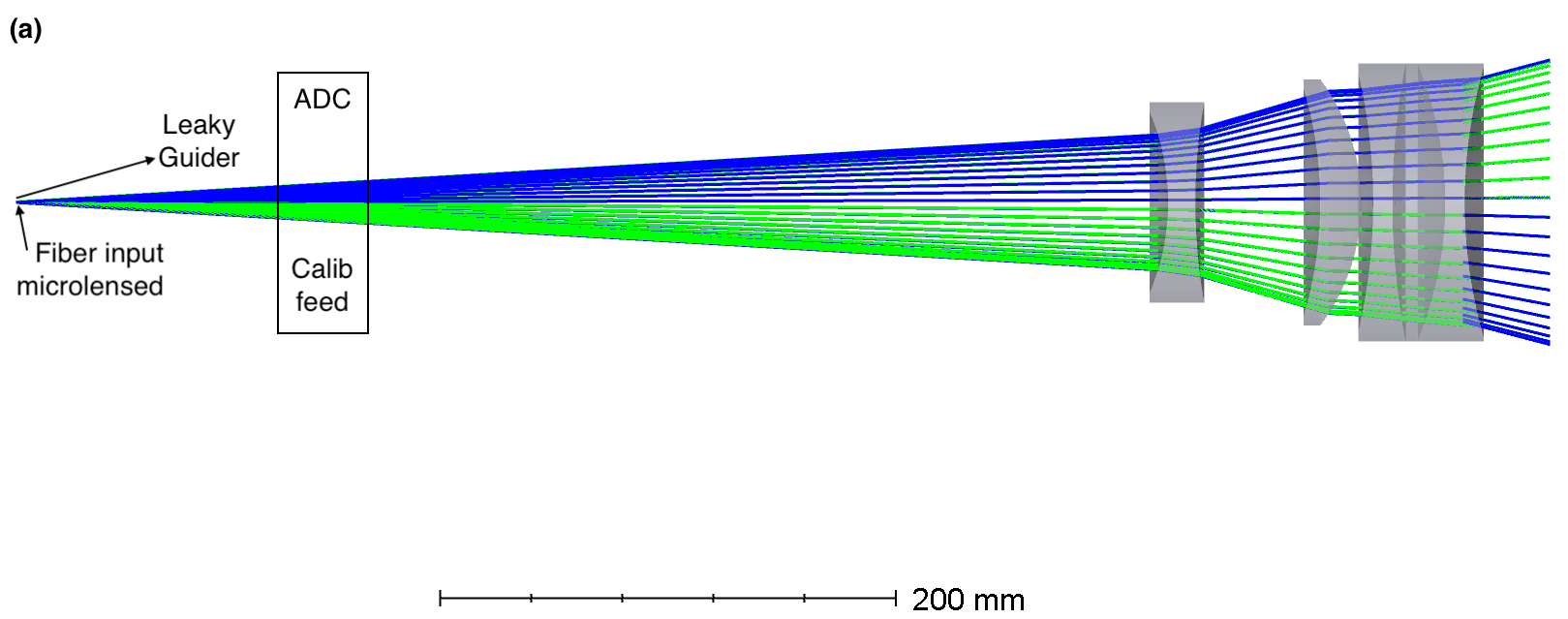}
\includegraphics[width=\textwidth]{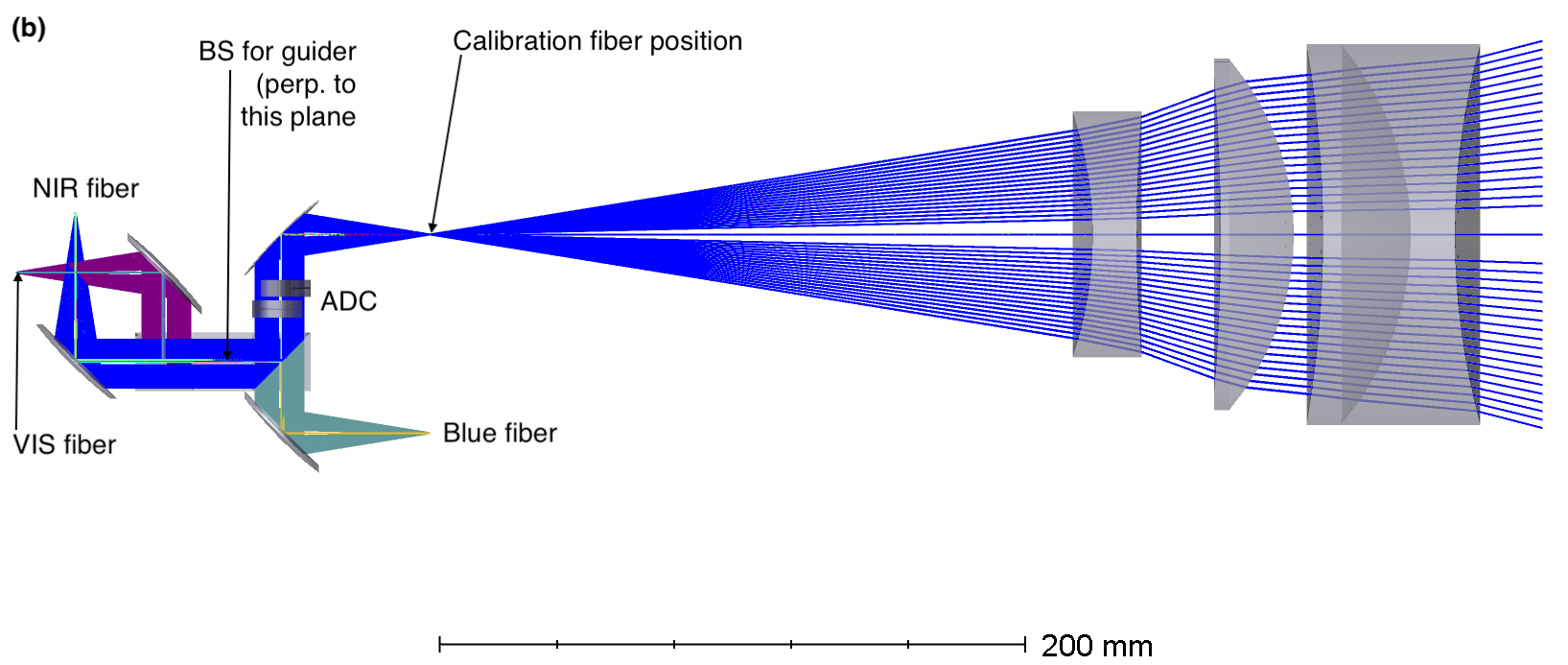}
\caption{\label{fig:corrector}
Optical layout of the STELLA-II current and new prime focus correctors. The main mirror is located to the right. 
{\em a:} The current corrector creates a f/8 beam and therefore creates a very long focal unit. The telescope currently clears the inner roof structure only by a few cm.
{\em b:} The new corrector design creates a f/3 beam and uses off-axis parabolas to collimate the beam for the ADC and to fold the three light paths for the three fibers. Despite the additional beam-splitters and extra fiber feeds, the total length of the system is smaller than the old one.
}
\end{figure}

Special care has been taken in the design and especially the glass selection of the ADC. Usually, small offsets of the corrected beam can be tolerated and compensated during acquisition by a small telescope offset. But since we feed all three bandpasses simultaneously into their respective fibers, the ADC has to correct perfectly across the whole wavelength range. Furthermore, the transmission in the UV needs to be good for all the glasses used in the system (Fig.~\ref{fig:adc}). 

The new corrector design (Fig.~\ref{fig:corrector}b) will be more compact than the old design by using faster (f/3) beams and direct fiber coupling. We use a 90\degr off-axis parabolic (OAP) mirror to collimate the beam for the ADC and the two dichroic beam splitters (BS). We  feed the blue spectrograph after the blue dichroic BS to minimize glass losses, and add an additional grey BS before the red dichroic BS which will be used for guiding and control of the fiber illumination. The fibers will be illuminated directly by the same OAPs mounted symmetrically with respect to the collimating OAP, similar to the PEPSI-VATT focal station\cite{2016SPIE.9912E..5HS}.

The calibration light will be fed directly into the focus of the first OAP mirror, using a 200\,$\mu$m fiber from a calibration unit providing both flat field and ThAr light. All fibers will be of the same type, the Polymicro FBP fiber with 50\,$\mu$m core diameter and octagonal cross-section. At wavelengths of 380\,nm, the FBP fiber has still comparable throughput to more specialized UV fibers, but offers better understood FRD and scrambling properties.

\section{SES-H\&K: THE NEW BLUE SPECTROGRAPH}
\label{sec:seshk}

\begin{figure}[ht]
\centering
\includegraphics[width=\textwidth]{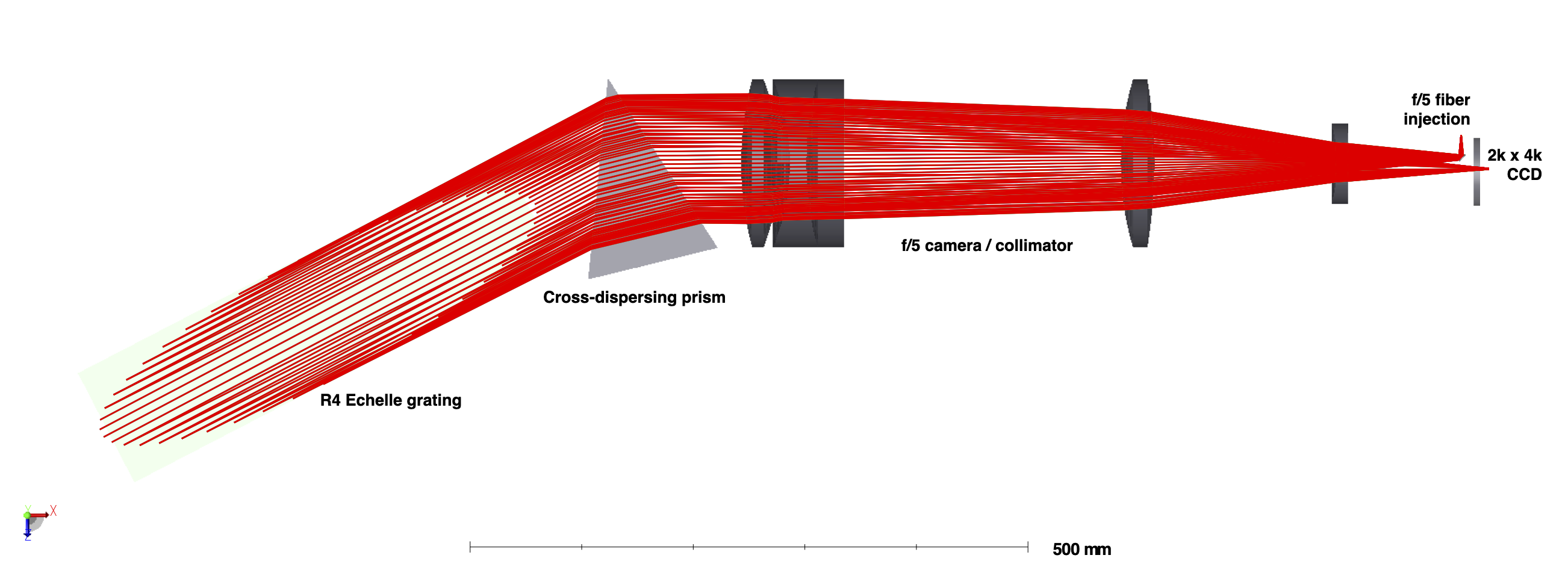}
\caption{\label{fig:seshuk}
Optical layout of the SES-H\&K spectrograph. As it will be located at the telescope, we use a double-pass design to keep the footprint small. To the left is the 110x420mm R4-\'Echelle grating, the cross-disperser is a PBM2Y-prism, and for a detector we will use a 4k STA CCD.
}
\end{figure}

\begin{figure}[ht]
\begin{minipage}[c]{.45\textwidth}
\includegraphics[width=\textwidth]{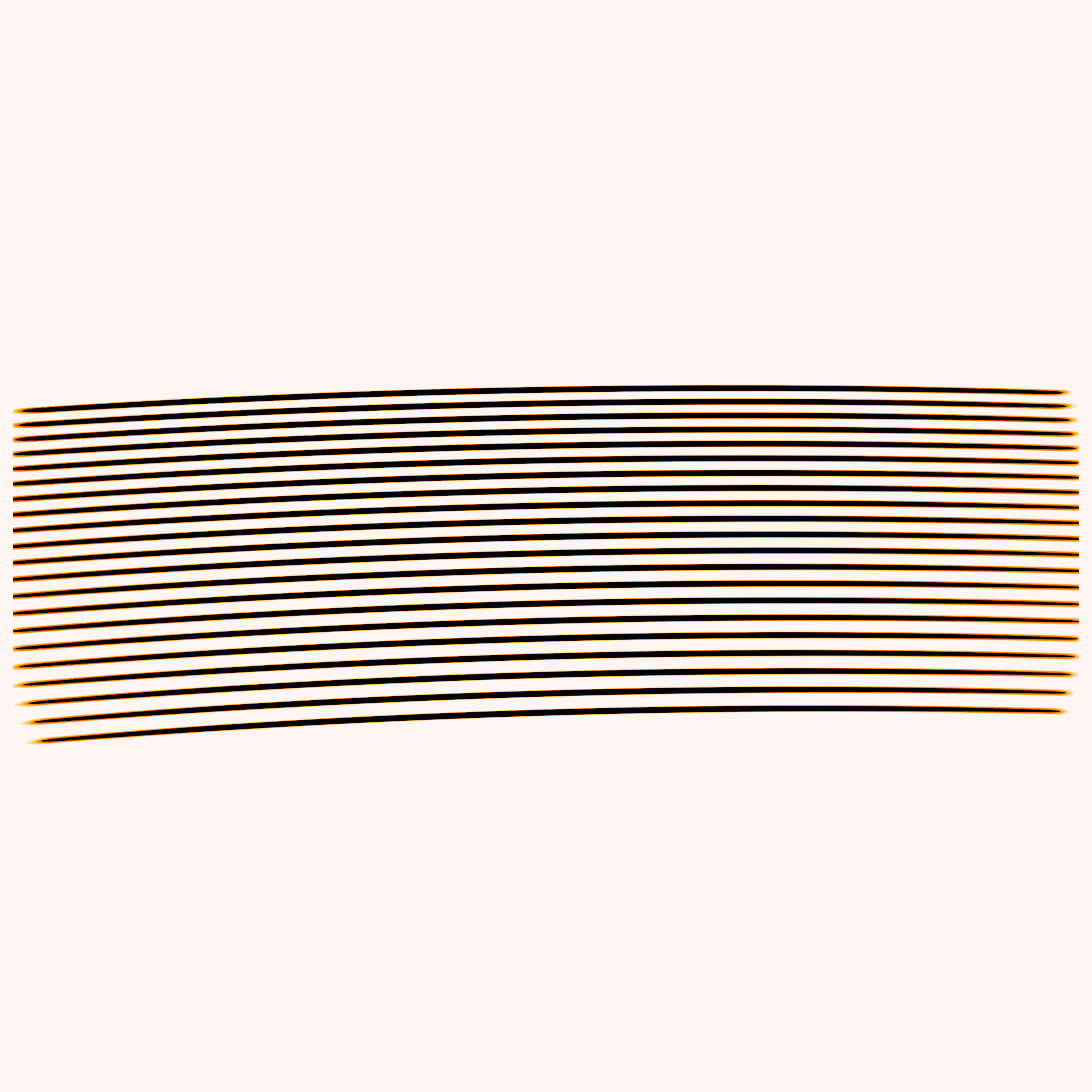}
\end{minipage}\hfill
\begin{minipage}[t]{.5\textwidth}
\caption{\label{fig:seshuk:flat}
Simulated flat-field image of SES-HK. The area covered by the spectrum is roughly 60x20\,mm, a perfect match for a 4k\,x\,2k CCD with 15\,$\mu$m pixels. But since the supply of square 4k detectors is much better, such a detector is foreseen.}
\end{minipage}
\end{figure}

The original SES spectrograph had a few legacy components with low or unknown blue throughput. After we replaced all of them during the upgrade when we moved to STELLA-II we noticed that even though efficiency has improved, the effect of the long fiber with a loss of approximately 25\% at 390\,nm is dominating the efficiency budget. Therefore, we decided to split the blue wavelength range off and make a dedicated blue optimized instrument, move it close to the telescope and compromise on stability and maybe resolution rather than on blue throughput. Fiber transmission losses at this distance will be smaller than 5\%. 

As the blue spectrograph should be small enough to fit on the telescope, we  opted for a double-pass design (see Fig.~\ref{fig:seshuk}). Using an existing R4-\'Echelle grating, we found that a f/5 design is a good trade-off between optical quality and system size. As a detector, a 4k CCD with 15$\mu$m pixels will be used. The slow f-number will allow for a very comfortable pixel sampling, while still guaranteeing full wavelength coverage without gaps. 

\begin{figure}[ht]
\centering
\includegraphics[width=0.5\textwidth]{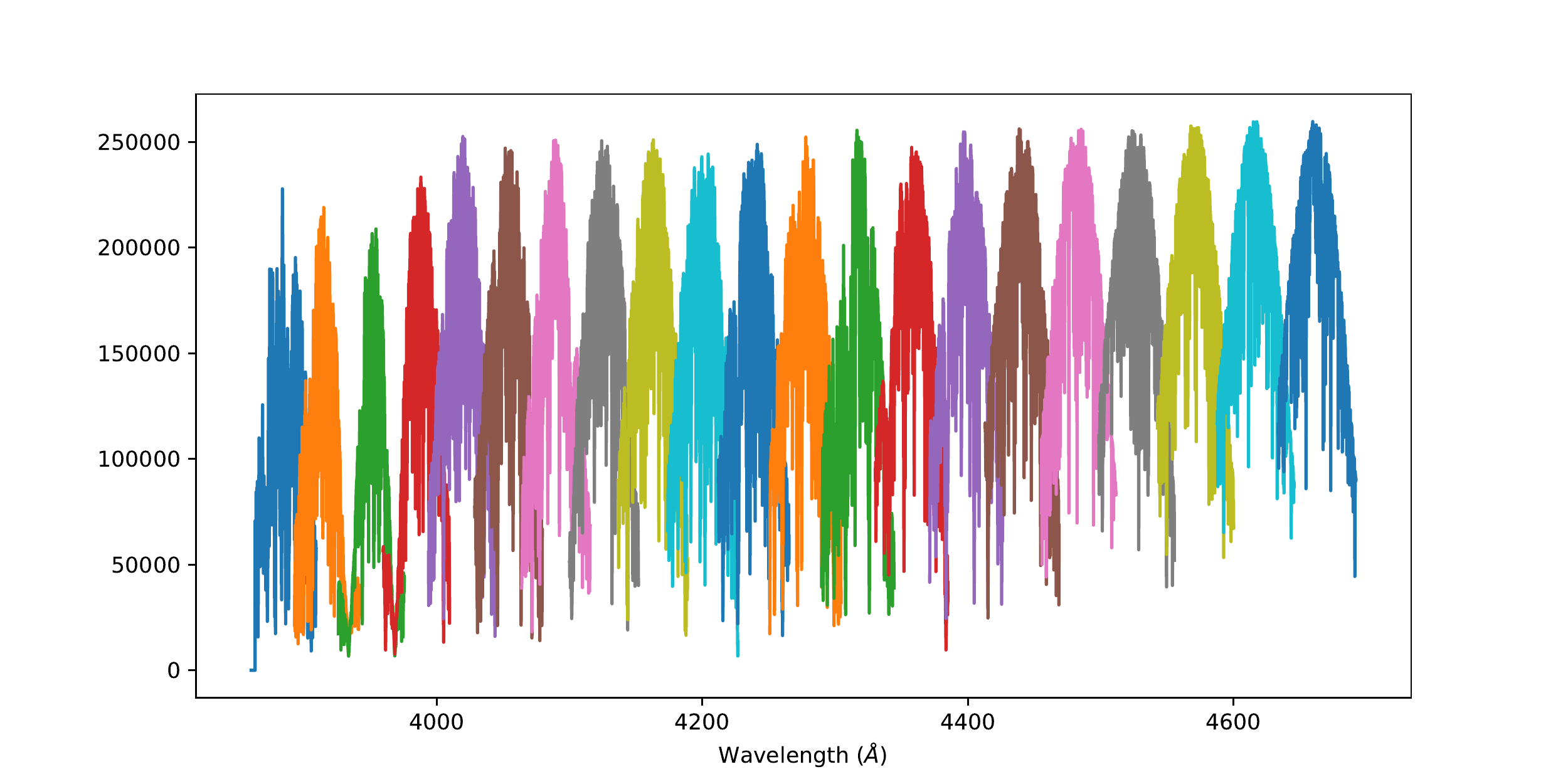}%
\includegraphics[width=0.5\textwidth]{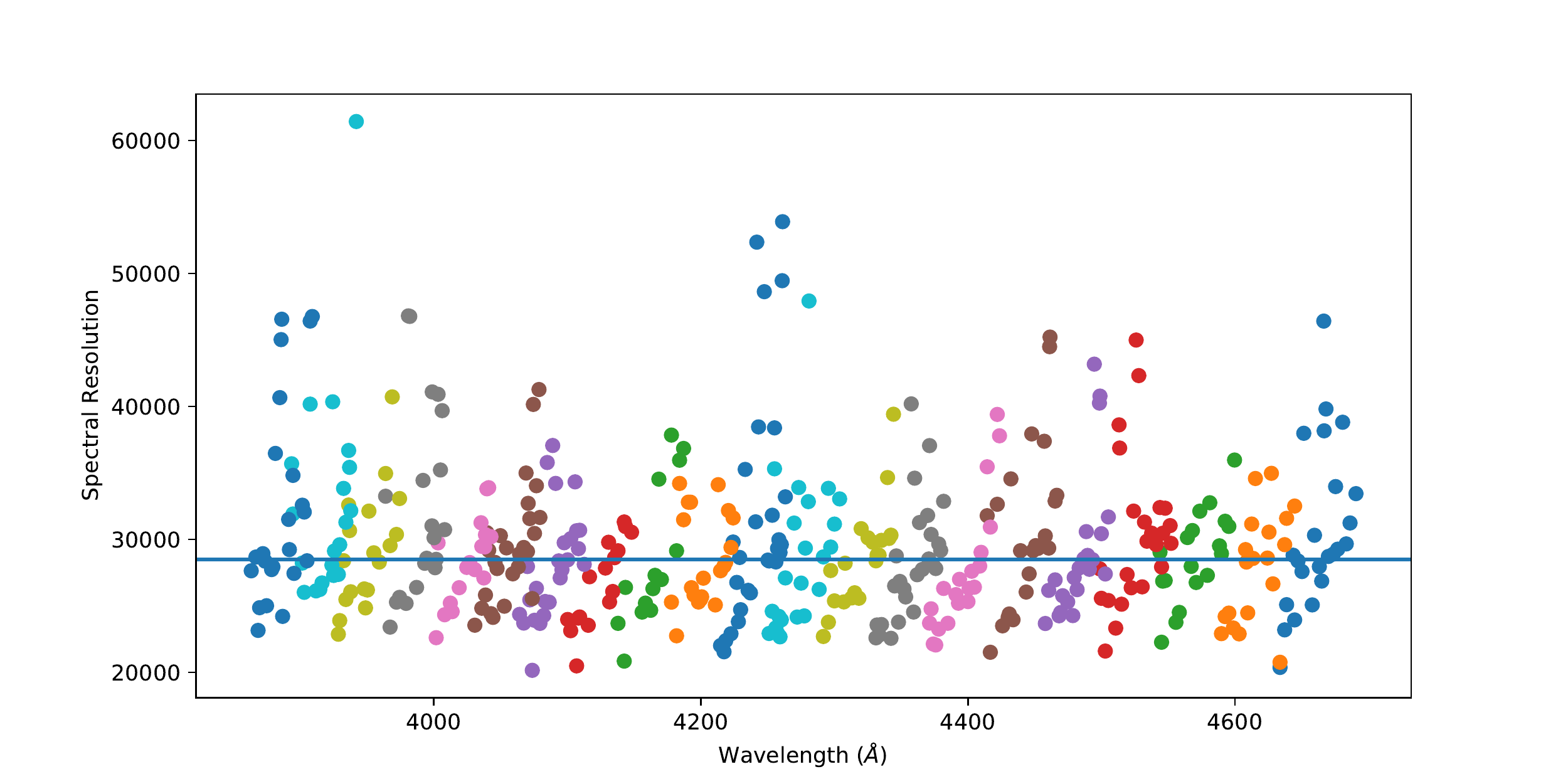}%
\caption{\label{fig:seshuk:spec}
{\em Left:} Simulated SES-H\&K spectrum of a solar-type star, the continuum of the flux-distribution is normalized, but the blaze function is not corrected for illustration purposes. 
{\em Right:} Resolution map derived from a simulated ThAr spectrum, the spectral orders are color coded. The median spectral resolution of 28\,000 is marked with a horizontal line.
}
\end{figure}

The wavelength range of the spectrograph will be from about 380\,nm in the blue to the blue cutoff of SES-VIS at 470\,nm. We use the 41.6 lines/mm R4 grating at its blaze angle of 76\degr and use 21 spectral orders from 100 to 120 covering a wavelength range of 386\,nm to 469\,nm. 
We used Optics Studio to simulate the expected spectra and used our spectral reduction package SDS4PEPSI\cite{2018A&A...612A..44S} to derive reduced spectra from a flat field, a ThAr-calibration lamp, and a solar type star. These spectra show the distribution of wavelength and flux over the spectral orders, and allow us to derive the effective spectral resolution from the width of the spectral lines (see Fig.~\ref{fig:seshuk:spec}).  The wavelength range is well covered, there is ample overlap between the spectral orders to allow for correction of differential illumination effects of science vs.\ calibration spectra. The sampling of the spectral resolution is about 6 pixels, and would allow for the implementation of an image slicer to double the resolving power of the spectrograph. Alternatively, we could bin the detector by two to reduce the read-noise.

\section{SES-VIS: THE STABILIZED VISUAL SPECTROGRAPH}
\label{sec:sesvis}

SES-VIS  (previously known as  GREGOR At Night Spectrograph GANS\cite{2018SPIE10702E..6LJ}) uses a classical white pupil optical design. Its design goals were to measure precise radial velocities, it is therefore contained inside a vacuum vessel for stability. We use a R4-\'Echelle grating, with orders 89 - 130 for wavelengths from 468\,nm to 691\,nm. More details can be found in the presentation from SPIE 2018\cite{2018SPIE10705E..1EW}.  

\begin{figure}[ht]
\begin{minipage}[c]{.45\textwidth}
\includegraphics[width=\textwidth]{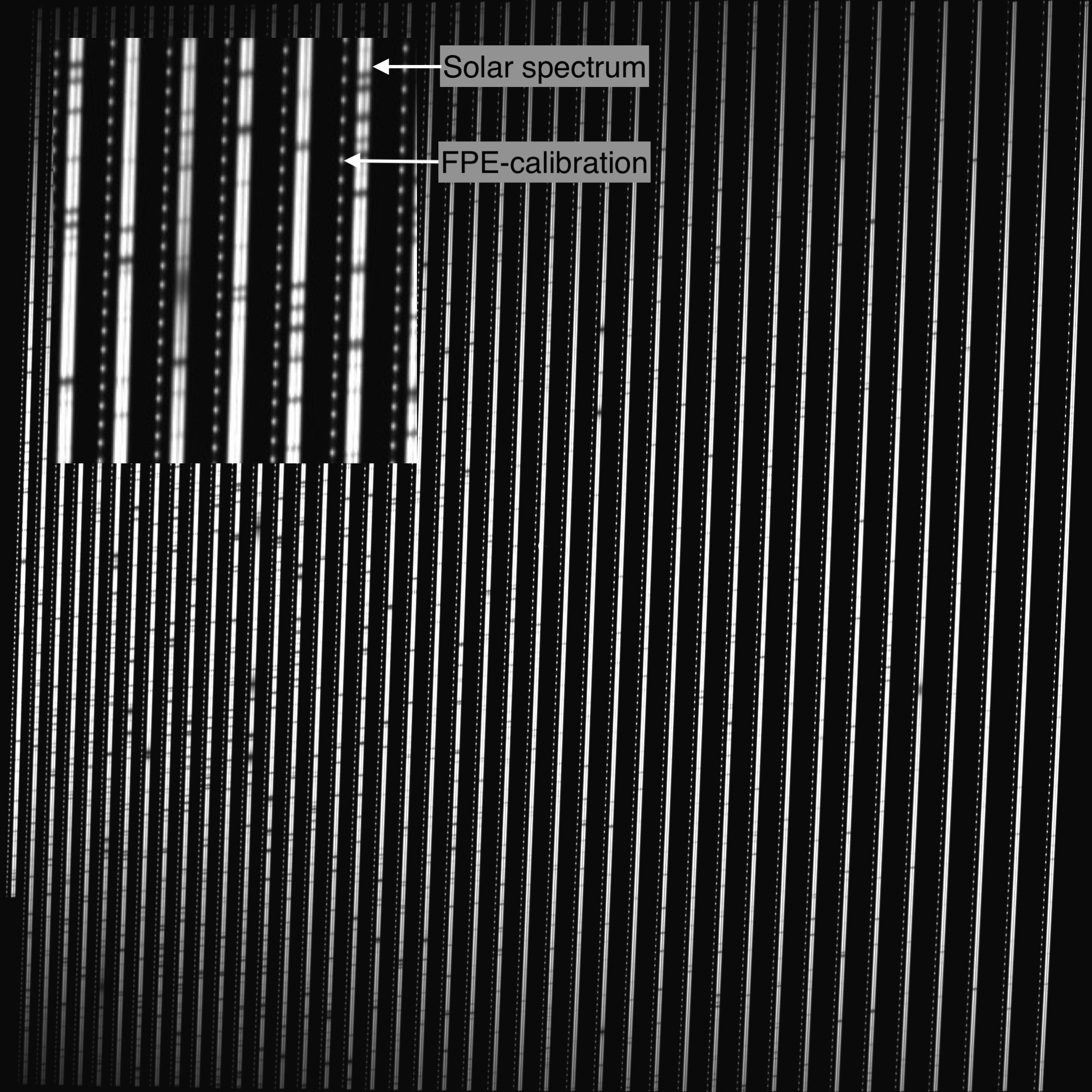}
\end{minipage}\hfill
\begin{minipage}[t]{.5\textwidth}
\caption{\label{fig:sesvis:sunfp}
Simulated science exposure of SES-VIS. We used a solar-type model spectrum through a double image-slicer. Next to each science order (at the left side) is its corresponding FPE-calibration spectrum, more clearly visible in the insert.}
\end{minipage}
\end{figure}

As with SES-H\&K above, we used the optical design to create simulated flat fields, ThAr calibration images and science images. In order to monitor the stability of the spectrograph, a Fabry-Perot etalon (FPE) is used to create a stable line-pattern between the spectral orders simultaneously to the science or calibration light. The position of these FPE-lines are calibrated periodically throughout the night,  and the appropriate shifts recorded together with the meta-data of the corresponding observation. An example solar spectrum is shown in figure~\ref{fig:sesvis:sunfp}, with the FPE-spectrum recorded on the side of each spectral order of the science observation.

\begin{figure}[hb]
\includegraphics[width=0.5\textwidth]{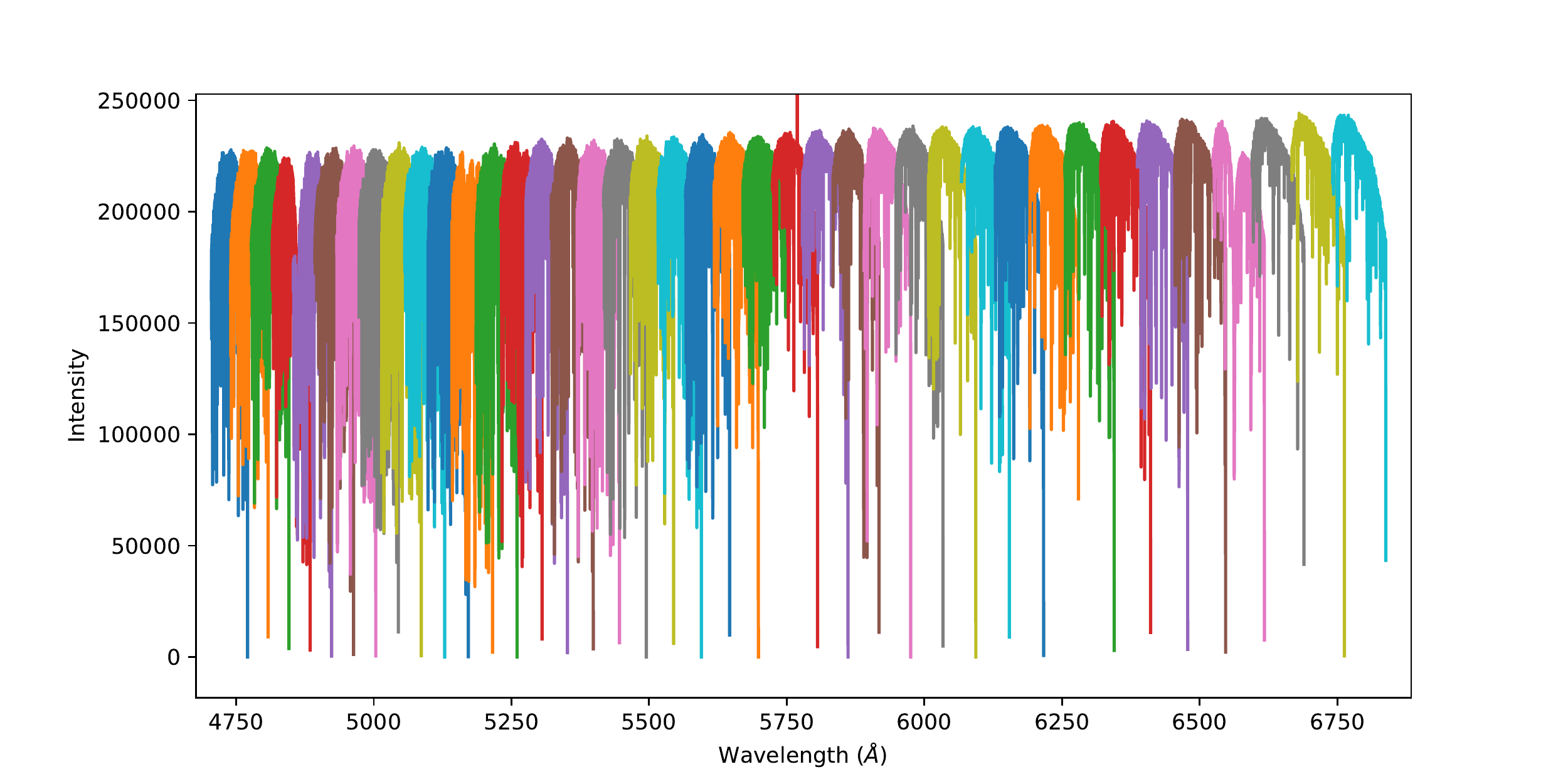}
\includegraphics[width=0.5\textwidth]{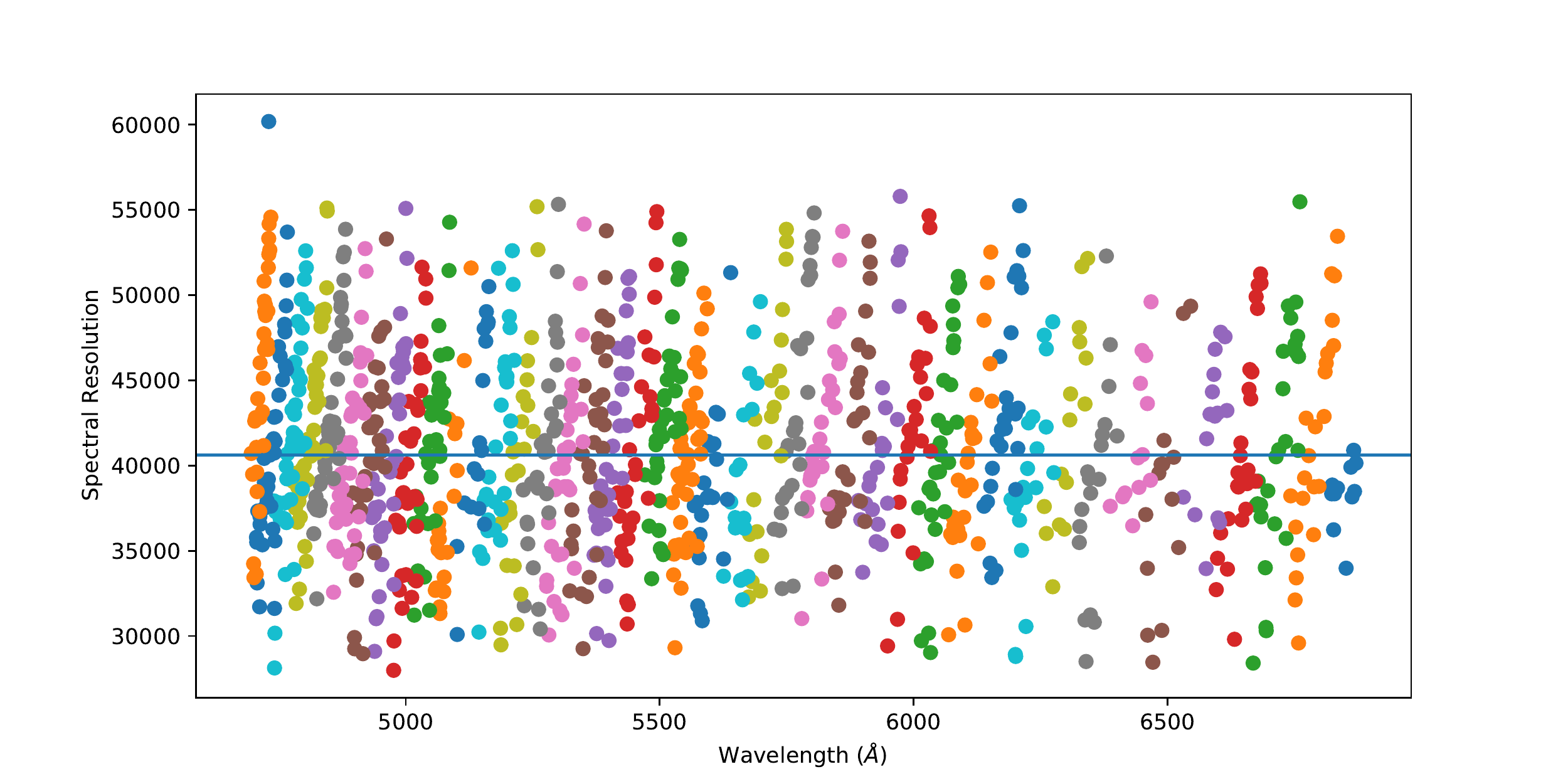}
\caption{\label{fig:sesvis:sun}
Simulated SES-VIS spectra derived from the optical model. 
{\em Left:} Spectrum of a solar like star with normalized intensity, the H$\alpha$-line can be seen in the 4th-last order. Note that the blaze function is not corrected here for illustration purposes.
{\em Right:} Resolution map derived from a simulated ThAr spectrum, the spectral orders are color coded, the median spectral resolution of 40\,000 is marked with a horizontal line.
}
\end{figure}

A simulated spectrum of a solar-type star with constant continuum is shown in Fig.~\ref{fig:sesvis:sun}, with the intensity variations due to the blaze function of the Échelle-grating still imprinted. The overlap between the orders is sufficient throughout the spectral range to aid in correcting possible continuum differences between the calibration flat field image and the science images. Also shown in Fig.~\ref{fig:sesvis:sun} to the right is the spectral resolution map derived from a simulated ThAr calibration image. The spectral resolution is derived from the line width of the ThAr-lines, shown are the widths of the detected ThAr-lines, color coded by spectral order to make it easier to see the effect of the anamorphic magnification. The median spectral resolution is 40\,000.

The thermal enclosure and the vacuum tank has undergone extensive tests in the previous years, and the bulky parts have been packed for shipping to Tenerife. As we did not want to set up the spectrograph in the dirty integration hall area, the spectrograph itself is going to be aligned and tested in the lab lab until the shipping to Tenerife in mid 2021.

\section{SES-NIR: THE CURRENT STELLA ECHELLE SPECTROGRAPH}
\label{sec:sesnir}

Currently, SES\cite{2008SPIE.7019E..19W} records 82 spectral orders (numbers 65 to 146, R2 grating) covering wavelengths from 390\,nm to 880\,nm without gaps. The wavelength range was chosen to include both the H\&K-lines on the blue end as well as the Ca-triplet on the red end of the spectrum. After the efficiency upgrade almost 10 years ago, the components of the spectrograph are capable of reaching the cutoff-wavelength of the CCD (e2v, astro-broadband, deep depleted) at 1.03\,$\mu$m with 5\% efficiency remaining. However, we decided to keep the original wavelength coverage. Now, with all of the light up to 690\,nm being diverted to the new spectrographs, we will re-align the optical camera setup to move the NIR into the center of the CCD to speed up readout time. This will allow us to include the lower spectral orders up to about 1.05\,$\mu$m. The new spectral format will thus include 29 spectral orders (54 to 82), covering the wavelength range from 690\,nm to 1.05\,$\mu$m. 

Ideally, we would like to extend the wavelength range even further into the IR, but that is not possible using current CCD technology (our detector is already one of the most IR-sensitive CCD detectors on the market). We tried to source a small InGaAs detector to be able to cover the scientifically interesting 1.08\,$\mu$m He-triplet. A small change in the tilt of the grating would get that region onto the detector, while still preserving coverage of the Ca-triplet. Of course a complete coverage can not be achieved with the small detectors available. Unfortunately, we found no commercial detector system capable of the task. We are currently assessing if it is worthwhile to manufacture a prototype-detector system, and will proceed using the current detector in the meantime.

\acknowledgments % equivalent to \section*{ACKNOWLEDGMENTS}       
The STELLA facility is funded by the Science and Culture Ministry of the German State of Brandenburg (MWFK) and the German Federal Ministry for Education and Research (BMBF), and is operated by the AIP jointly with the IAC.

% References
\bibliography{aa_mnem,stellaspie} % bibliography data in report.bib
\bibliographystyle{spiebib} % makes bibtex use spiebib.bst

\end{document}